\documentclass[prd,amsmath,amssymb,showpacs,superscriptaddress,nofootinbib,twocolumn]{revtex4-1}
\usepackage{multirow}
\usepackage{float}
\usepackage{graphicx}
\usepackage{dcolumn}
\usepackage{bm}
\usepackage{rotating}
\usepackage{color}
\usepackage{subfigure}
\usepackage{pstricks}
\usepackage{soul}
\usepackage{overpic}
\usepackage[english]{babel}
\usepackage[colorlinks,urlcolor=black,linkcolor=blue,anchorcolor=blue,citecolor=blue]{hyperref}
\usepackage{tabularx}
\usepackage{lineno}
\newcommand{\PreserveBackslash}[1]{\let\temp=\\#1\let\\=\temp}
\newcolumntype{C}[1]{>{\PreserveBackslash\centering}p{#1}}
\newcolumntype{R}[1]{>{\PreserveBackslash\raggedleft}p{#1}}
\newcolumntype{L}[1]{>{\PreserveBackslash\raggedright}p{#1}}

\usepackage{lipsum} 
\newlength{\halfpagewidth}
\newcommand{\tabincell}[2]{\begin{tabular}{@{}#1@{}}#2\end{tabular}}
\newcommand{\Rone}{\uppercase\expandafter{\romannumeral1}}
\newcommand{\Rtwo}{\uppercase\expandafter{\romannumeral2}}
\newcommand{\Rthree}{\uppercase\expandafter{\romannumeral3}}
\newcommand{\Rfour}{\uppercase\expandafter{\romannumeral4}}
\newcommand{\threes}{~~~}
\newcommand{\jpsi}{J/\psi}

\begin{document}
%
\title{Some remarks on $X(6900)$}
\author{Qin-Fang Cao}
\affiliation{Department of Physics and State Key Laboratory of Nuclear Physics and Technology,
Peking University, Beijing 100871, China}
\author{Hao Chen}
\affiliation{Department of Physics and State Key Laboratory of Nuclear Physics and Technology,
Peking University, Beijing 100871, China}
\author{Hong-Rong Qi}
\thanks{Corresponding author. qihongrong@tsinghua.edu.cn}
\affiliation{Department of Engineering Physics, Tsinghua University, Beijing 100084, China}
\author{Han-Qing Zheng}
\email{Other e-mail addresses: caoqf@pku.edu.cn (Qin-Fang Cao), haochen0393@pku.edu.cn (Hao Chen),  zhenghq@pku.edu.cn (Han-Qing Zheng).}
\affiliation{Department of Physics and State Key Laboratory of Nuclear Physics and Technology,
Peking University, Beijing 100871, China}
\affiliation{Collaborative Innovation Center of Quantum Matter, Beijing 100871, China}

\date{\today}

\begin{abstract}
The analysis of the  LHCb data on $X(6900)$ found in the di-$\jpsi$ system
is performed using a momentum-dependent  Flatt\'{e}-like parameterization.
The use of the pole counting rule and spectral density function sum rule give consistent conclusions that both confining states and  molecular states are possible, or it is unable  to distinguish the nature of $X(6900)$, if only the di-$\jpsi$ experimental data with current statistics are available.
Nevertheless, we found that the lowest state in the di-$J/\psi$ system has very likely the same quantum numbers as $X(6900)$, and $X(6900)$ is probably not interpreted as a $J/\psi-\psi(2S)$ molecular state.
\end{abstract}

\maketitle

\section{Introduction}
\label{intro}

Recently, LHCb Collaboration observed a structure around 6900 MeV/$c^2$, dubbed as $X(6900)$, in the di-$\jpsi$ invariant mass spectrum~\cite{6900}, with the signal statistical significance of above 5$~\sigma$. It is probably composed of four (anti)charm quarks ($c\bar{c} c\bar{c}$) and its width~\cite{6900} are determined to be $80\pm19~({\rm stat.})\pm33~({\rm sys.})$ and $168\pm33~({\rm stat.})\pm69~({\rm sys.})$ MeV in two fitting scenarios of  Breit-Wigner parameterizations with the constant widths. Additionally,  a broad bump and a narrow bump exist in the low and high sides of the di-$\jpsi$ mass~\cite{6900}, respectively, where the former might be a result from a lower broad resonant state (or several lower states)  or interference effect, and the latter is found to be a hint of a state located at $\sim$7200 MeV, called $X(7200)$.

The intriguing observation has aroused widespread concern in  physics community. In accordance with QCD sum rule,
Ref.~\cite{HXChen} pointed out that the lowest broad structure between 6200 and 6800 MeV can be regarded as an $S$-wave $cc\bar{c}\bar{c}$ tetraquark state with  the quantum numbers $J^{PC}=0^{++}$ or $2^{++}$, while the $X(6900)$ as a $P$-wave $cc\bar{c}\bar{c}$ tetraquark with $J^{PC}=0^{-+}$ or $1^{-+}$.
In the framework of a non-relativistic potential quark model (NRPQM) for heavy quark system, Ref.~\cite{MSLiu} deemed that the lowest one can be interpreted by an $S$-wave state around 6500 MeV, and the $X(6900)$ by a $P$-wave $cc\bar{c}\bar{c}$ state.
Also in NRPQM, Ref.~\cite{Wang:2021kfv} takes $X(6900)$ as a candidate of the first radially excited tetraquarks with $J^{PC}=0^{++}$ or $2^{++}$, or the $1^{+-}$ or $2^{-+}$ $P$-wave state, and considered that there exist two states below $X(6900)$, which have exotic quantum numbers $0^{--}$ and $1^{-+}$ and may decay into the $P-$wave $\eta_c J/\psi$ and di-$\jpsi$ modes, respectively.
Ref.~\cite{QFLv} indicated, in an extended relativistic quark model,  that the lowest broad structure should contain one or more ground $cc\bar{c}\bar{c}$ tetraquark states, while the narrow structure near 6900 MeV can be categorized as the first radial excitation of $cc\bar{c}\bar{c}$ system.
Exploiting three potential models (a color-magnetic interaction model, a traditional constituent quark model, and a multiquark color flux-tube model), Ref.~\cite{Deng:2020iqw} systematically investigated the properties of the states $[Q_1Q_2][\bar{Q}_3\bar{Q}_4]~(Q=c,b)$: the broad structure ranging from 6200 to 6800 MeV can be described as the ground tetraquark state $[c\bar{c}][c\bar{c}]$ in the three models, while the narrow $X(6900)$ exhibits different properties in different potential models and more data associated with determination of quantum numbers are needed to shed light on the nature of these states.
Ref.~\cite{Dong:2020nwy} argued that the $X(6900)$ structure can be well described within two variants of a unitary couple-channel approach: (i) with two channels $\jpsi\jpsi$ and $\jpsi\psi(2S)$ with energy-dependent interactions, or (ii) with three channels $\jpsi\jpsi$, $\jpsi\psi(2S)$ and $\jpsi\psi(3770)$ with just constant contact interactions.
They predicted, moreover,  the existence of a near-threshold state $X(6200)$~\cite{Dong:2020nwy} in the $\jpsi\jpsi$ system with the quantum numbers $J^{PC} = 0^{++}$ or $2^{++}$.
Similarly, in coupled-channel analyses, Ref.~\cite{DLYao} identified $X(6900)$ as $2^{++}$, and provided hints of the existence of the other states: a $0^{++}$  $X(6200)$, a $2^{++}$ $X(6680)$, and a $0^{++}$ $X(7200)$, and Ref.~\cite{ZHGuo} predicted a narrow resonance $X(6825)$ located below the $\chi_{c0}\chi_{c0}$ threshold and of molecular origin.
Employing a contact-interaction effective field theory with heavy anti-quark di-quark symmetry, Ref.~\cite{LSGeng} implied that $X(7200)$ can be regarded as the fully heavy quark partner of $X(3872)$.
Ref.~\cite{ZhaoQ} showed that the structure $X(6900)$, as a dynamically generated resonance pole, can arise from Pomeron exchanges and coupled-channel effects between the $\jpsi\jpsi$, $\jpsi\psi(2S)$ scatterings.
Based on perturbative QCD method, Ref.~\cite{YQMa} found that there should exist another state near the resonance at around 6.9 GeV, and the ratio of production cross sections of $X(6900)$ to the undiscovered state is very sensitive to the nature of $X(6900)$.
Besides discussing the nature of $X(6900)$, Ref.~\cite{XYWang} studied the production of $X(6900)$ in $p\bar{p}\to \jpsi\jpsi$ reaction within an effective Lagrangian approach and Breit-Wigner formula, and predicted that it is feasible to find $X(6900)$ in the $p\bar{p}$ collision in D0 and forthcoming PADNA experiments.

Generally, a molecular state may locate near the threshold of two (or more) color singlet hadrons, like deuteron, $Z_b(10610)$~\cite{BelleZb}, $Z_c(3900)$~\cite{BESIII3900,Belle3900,CLOE3900}, $P_c(4470)$~\cite{Pc2015,Pc2019}, $etc.$
We found that the $X(6900)$ state is close to the threshold of $\jpsi \psi(3770)$, $\jpsi \psi_2(3823)$, $\jpsi$$\psi_3(3842)$, and $\chi_{c0} \chi_{c1}$; 
and the $X(7200)$ is close to the threshold of  $\jpsi \psi(4160)$ and $\chi_{c0}$$\chi_{c1}(3872)$. 
Inspired by this, in this paper, based on the assumption of $X(6900)$ coupling to $\jpsi\jpsi$, $\jpsi\psi(3770)$, $\jpsi \psi_2(3823)$, $\jpsi \psi_3(3842)$ and $\chi_{c0}\chi_{c1}$ processes (see Tab.~\ref{tab:Swave}), and  $X(7200)$ to  $\jpsi\jpsi$, $\jpsi\psi(4160)$ and $\chi_{c0}\chi_{c1}(3872)$ (see Tab.~\ref{tab:Pwave}, where parameters of charmonia used in the analysis see Tab.~\ref{tab:particles} for details),  a  Flatt\'{e}-like parameterization with momentum-dependent partial widths for the two resonances is used to fit the experimental data, and then the pole positions of the scattering amplitude in the complex $s$ plane are searched for.
For the $S$-wave $\jpsi \jpsi$ coupling, the pole counting rule (PCR)~\cite{pole}, which has been applied to the studies of ``$XYZ$" physics in Refs.~\cite{Zhang:2009bv,Dai:2012pb,X3900,Cao:2019wwt}, and spectral density function sum rule (SDFSR)~\cite{X3900,Baru:2003qq,Weinberg,Weinberg:1965zz,Kalashnikova:2009gt}   are employed to analyze the nature of the two structures, i.e., whether they are more inclined to be confining states bound by color force, or  loosely-bounded hadronic molecular states.
\begin{table*}[htbp]
    \centering
    \caption{Involved $S$-wave couple channels except di-$J/\psi$.}
    \label{tab:Swave}
    \begin{tabular*}{\textwidth}{@{\extracolsep{\fill}}ccccc}
    \hline
    \hline\noalign{\smallskip}
      $J^{PC}\text{of di-}J/\psi$   & Couple channels of $X(6900)$  &  Threshold (MeV) & Couple channels of $X(7200)$  &  Threshold (MeV)  \\
      \noalign{\smallskip}\hline\noalign{\smallskip}
      $0^{++}$         & \tabincell{c}{ $\jpsi-\psi(2S)$ \\  $J/\psi-\psi(3770)$ }      &  \tabincell{c}{ 6783.0 \\ 6870.6 }    & $J/\psi-\psi(4160)$                   &  7287.9 \\
      \noalign{\smallskip}\hline\noalign{\smallskip}
      $2^{++}$ &\tabincell{c}{ $\jpsi-\psi(2S)$ \\  $J/\psi-\psi(3770)$\\  $J/\psi-\psi_2(3823)$\\ $J/\psi-\psi_3(3842)$} &\tabincell{c}{ 6783.0\\ 6870.6\\  6919.1\\ 6939.6}  &  $J/\psi-\psi(4160)$   &  7287.9\\
      \noalign{\smallskip}\hline
      \hline
    \end{tabular*}
\end{table*}
\begin{table*}[htbp]
    \centering
    \caption{Involved $P$-wave couple channels except di-$J/\psi$.}
    \label{tab:Pwave}
    \begin{tabular*}{\textwidth}{@{\extracolsep{\fill}}ccccc}
    \hline
    \hline\noalign{\smallskip}
      $J^{PC}\text{of di-}J/\psi$   & Couple channels of $X(6900)$       &  Threshold (MeV)  & Couple channels of $X(7200)$  &  Threshold (MeV)  \\
      \noalign{\smallskip}\hline\noalign{\smallskip}
      \tabincell{c}{ $1^{-+}$ \\ $(0,1,2)^{-+}$ }  &\tabincell{c}{$\chi_{c0}-\chi_{c1}$\\ $J/\psi-\psi(3770)$} & \tabincell{c}{ 6925.4\\ 6870.6}  & \tabincell{c}{ $\chi_{c0}-\chi_{c1}(3872)$\\ $J/\psi-\psi(4160)$} &\tabincell{c}{ 7286.4\\ 7287.9}  \\
      \noalign{\smallskip}\hline
      \hline
    \end{tabular*}
\end{table*}
\begin{table*}[htpb]
    \centering
    \caption{Parameters for the involved charmonium states~\cite{PDG}.}
    \label{tab:particles}
    \begin{tabular*}{\textwidth}{@{\extracolsep{\fill}}cccccccccc}
    \hline
    \hline\noalign{\smallskip}
                               & $J/\psi$      &   $\chi_{c0}$   &  $\chi_{c1}$  &  $\psi(2S)$ & $\psi(3770)$  &  $\psi_2(3823)$ & $\psi_3(3842)$  &  $\chi_{c1}(3872)$  & $\psi(4160)$   \\
         \noalign{\smallskip}\hline\noalign{\smallskip}
       $J^{PC}$           &  $1^{--}$     &    $0^{++}$     &   $1^{++}$     &  $1^{--}$       &     $1^{--}$      &   $2^{--}$          &   $3^{--}$     &    $1^{++}$             &    $1^{--}$       \\
       mass (MeV)        & 3096.9        &      3414.7      &    3510.7       &   3686.1         &    3773.7          &   3822.2             &    3842.7       &    3871.7                &    4191.0          \\
       $n^{2S+1} L_{J}$ & $1^{3}S_{1}$ & $1^{3}P_{0}$   &  $1^{3}P_{1}$ &   $2^{3}S_1$   &    $1^{3}D_{1}$  &   $1^{3}D_2$       &   $1^{3}D_3$   &  $2^{3}P_{1}$~\cite{Meng:2014ota}         &  $2^{3}D_{1}$   \\
       \noalign{\smallskip}\hline
       \hline
    \end{tabular*}
\end{table*}
We also discussed the $X(6900)\to \jpsi\psi(2S)$ coupling with the threshold  below $X(6900)$'s mass of $\sim 100$ MeV.
This threshold is far away from the mass of $X(6900)$ so that it seems not like a $\jpsi-\psi(2S)$ molecular state,
but the process is easily accessible in experiments.


\section{Parameterization and pole counting}
\label{sec:2}
The states of $X(6900)$ and $X(7200)$ are parameterized with a momentum-dependent Flatt\'{e}-like formula. The non-resonance background shape is parametrized by the two-body phase space of $R\to \jpsi\jpsi$ times an exponential function. 
In order to better meet the di-$\jpsi$ spectrum,  a Flatt\'{e}-like function with only considering the $\jpsi\jpsi$ channel for the structure below 6800 MeV  is employed in the fit.
Not identifying the lowest state that contributes to the peak around 6500 MeV  or that corresponds to the dip (caused by destructive interference) below 6800 MeV, 
we do not analyze the nature of the lowest state (named as $X(6500)$ hereafter).
If excluding $X(6500)$ in the fit, it turns out not to be converged.
It means that a state with the same quantum numbers as $X(6900)$ is essential to describe the extremely deep dip below 6800 MeV by destructive interference.
As mentioned above, the components of the fit can be written,
\begin{equation}\label{components}
    \begin{aligned}
       &\mathcal{M}_{1}=\frac{g_{1} n_{11}(s) e^{i \phi_{1}}}{s-M_{1}^{2}+i M_{1} \Gamma_{11}(s)}, \\
        &\mathcal{M}_{i}=\frac{g_{i}n_{i1}(s) e^{i \phi_{i}}}{s-M_{i}^{2}+i M_{i} \sum_{j=1}^{2} \Gamma_{i j}(s)}, \\
       &\mathcal{M}_{\rm NoR}=c_0 e^{c_1(\sqrt{s}-2 m)} \sqrt{\frac{s-4 m^{2}}{s}}, 
\end{aligned}
\end{equation}
where  $M_1~(\Gamma_1)$ is the line-shape mass (width) for $X(6500)$, $m$ is the $\jpsi$ mass~\cite{PDG},
$M_i$  ($i=2,3$) corresponds to the line-shape  mass of $X(6900)$ and $X(7200)$, respectively;
$\Gamma_{ij}$ corresponds to  the partial width of the $j$-th couple channel on the $i$-th pole;
$\phi_1$ and $\phi_i$ are interference phases; $g_1$, $g_i$, $c_0$, and $c_1$ are free constants;
 $n_{ij}(s)$ combines the threshold and barrier factors; and $j=1$ represents the $\jpsi \jpsi$ channel (throughout the analysis).
The $n_{ij}(s)$ and $\Gamma_{ij}$ can be expressed~\cite{PDG},
\begin{equation}
\label{eq:Gamma}
    n_{ij}(s)=\left(\frac{p_{ij}}{p_0}\right)^l F_l(p_{ij}/p_0), \quad \Gamma_{ij}(s)=g_{ij}\rho_{ij}(s)n_{ij}^2(s),
\end{equation}
where, $l$ is the orbital angular momentum in channel $j$, $p_{ij}$ is the center-of-mass momentum of one daughter particle of channel $j$ for two body decays\footnote{For tow-body final states $m_a$ and $m_b$, $p=\sqrt{[s-(m_a+m_b)^2][s-(m_a-m_b)^2]/(4s)}$; for $p_j^2<0$,  $p_j$ is done using analytic continuation $p_j=i\sqrt{-p_j^2}$.}, $p_0$ denotes a momentum scale,
$g_{ij}$ is a coupling constant,  and $\rho_{ij}(s)=2p_{ij}/\sqrt{s}$, is the phase space factor.
The factor $p^l$ guarantees the correct threshold behavior. The rapid growth of this factor for angular momenta $l>0$ is commonly compensated at higher energies by the phenomenological form factor $F_l(p_{ij}/p_0)$. Often the Blatt-Weisskopf form factors are utilized~\cite{Blatt-W1,Blatt-W2,Blatt-W3}, $e.g.$ $F_{0}^{2}(z)=1, F_{1}^{2}(z)=1 /(1+z), F_{2}^{2}(z)=1 /\left(9+3 z+z^{2}\right)$ with $z=(p_{ij}/p_0)^2$.
Refs.~\cite{S.Kopp,2018ckj} give $z=(p_{ij}R)^2$, and they found that $R$, varying between 0.1 GeV$^{-1}$ and 10 GeV$^{-1}$, is a phenomenological factor (generally representing the ``radius" of a particle~\cite{S.Kopp}) with little sensitivity to the partial width. With $p_0$ and $R$ being positive real values, it is easy to find out $p_{ij}/p_0=p_{ij}R$. Therefore, $p_0$ varies between 0.1 GeV and 10 GeV, and is taken as 2 GeV in this analysis.

Due to limited data statistics, only two-channel couplings are investigated in the following:
A. $S-S$ couplings, B. $P-P$ couplings, C. $S-P$ and $P-S$ couplings,  where the former denotes the angular momentum of the $\jpsi\jpsi$ channel and the latter other channels listed in Tabs.~\ref{tab:Swave}-\ref{tab:Pwave}.
The corresponding pole positions of $X(6900)$ and $X(7200)$ are determined. 

\subsection{$S-S$ couplings}
\begin{figure}[htpb]
    \vskip 0.4cm
    \centering
    \subfigure[~Solution I]{\includegraphics[scale=0.22]{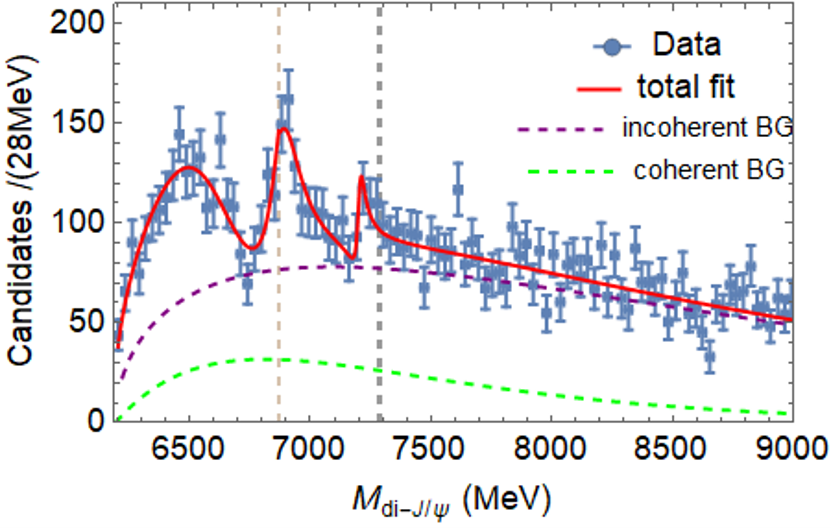}}
    \quad
    \subfigure[~Solution II]{\includegraphics[scale=0.22]{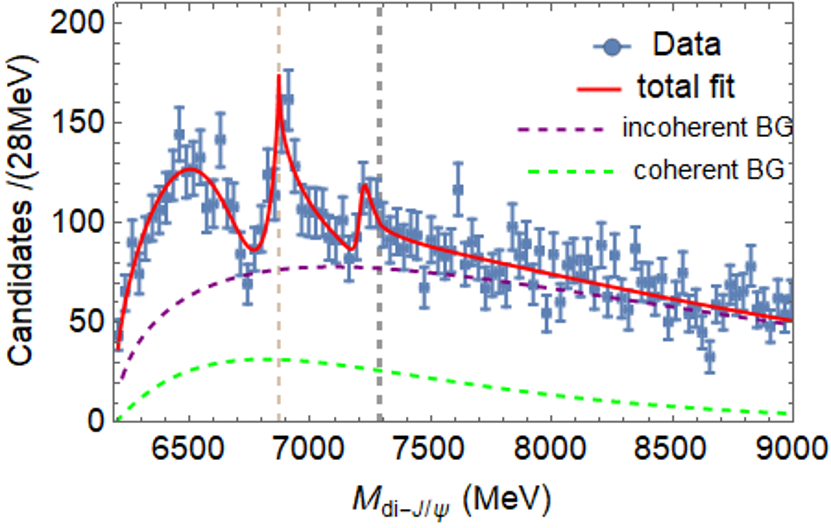}}
    \caption{(Color online) Fit projections of two solutions  for the $S-S$ couplings in Case I with the $X(6900)\to \jpsi\jpsi$ and $\jpsi \psi(3770)$, and $X(7200)\to \jpsi \jpsi$ and $\jpsi \psi(4160)$ couples, where the dots with error bars are the LHCb data~\cite{6900},  the red lines are the best fit, the green dashed lines show the coherent BG, the purple dashed lines are the contribution of the incoherent BG, and the vertical lines indicate the corresponding mass thresholds of $X(6900)$ and $X(7200)$.
}
\label{fig:sscouple1}
\end{figure}

Constrained by the generalized bose symmetry for identical particles and $J^{PC}$ conservation, the quantum numbers of the $S$-wave  $\jpsi \jpsi$ pair must be $0^{++}$ or $2^{++}$.
Based on the $0^{++}$ or $2^{++}$ assumption for $X(6900)$ and $X(7200)$, the other $S$-wave couple channels near the  mass of the two states are  considered, as summarized in Tab.~\ref{tab:Swave}.
They could be  divided into three cases for the $X(6900)$ decays:\\
\indent Case \uppercase\expandafter{\romannumeral1}: $\jpsi \jpsi$ and $\jpsi \psi(3770)$,     \\
\indent Case \uppercase\expandafter{\romannumeral2}: $\jpsi \jpsi$ and $\jpsi \psi_2(3823)$, \\
\indent Case \uppercase\expandafter{\romannumeral3}: $\jpsi \jpsi$ and $\jpsi \psi_3(3842)$. \\
For $X(7200)$, the $\jpsi \jpsi$ and the near-threshold $\jpsi$ $\psi(4160)$ channels are used in the couple channel analysis.
For the $X(6500)$ state, it has the same quantum numbers as $X(6900)$ (similarly hereafter), as has been noted.
Thus, the total amplitude $\mathcal{M}$ satisfies,
\begin{equation}\label{flatte}
    \begin{aligned}
      |\mathcal{M}|^2= &\Bigg| \sum_{i=1}^{3} \mathcal{M}_i + \mathcal{M}_{\rm NoR} \Bigg|^{2} + \mathcal{B.G.},
\end{aligned}
\end{equation}
where, $\mathcal{M}_{\rm NoR}$ describes the coherent background (BG), and incoherent background $\mathcal{B.G.}$  takes the similar parameterization as $\mathcal{M}_{\rm NoR}$. This background parameterization is similar to the LHCb experiment~\cite{6900}. 
Interestingly, two sets of solutions with almost equivalent goodness of the fit are found in all three cases, one of which favors that the two states are confining states and the other supports that they are molecular bound states, using both PCR and SDFSR mentioned above.
The fit results are summarized in Tab.~\ref{tab:swave:fit}. Since the fit curves of three cases look very similar, we only draw the fit projections of two solutions of case I in Fig.~\ref{fig:sscouple1}.

\begin{table*}[htpb]
    \centering
    \caption{Summary of numerical  results for  $S-S$ couplings, where $M_2$ and $M_3$ are the  mass of $X(6900)$ and $X(7200)$, respectively; $g_{21}$ and $g_{31}$ are the coupling constants of $X(6900)$ and $X(7200)$ decaying to $\jpsi \jpsi$, respectively; $g_{22}$ is the coupling constant of $X(6900)\to\jpsi \psi(3770)$, $\jpsi \psi_2(3823)$, and $\jpsi \psi_3(3842)$ in turn in the three cases, $g_{32}$ is the coupling constant of $X(7200)\to \jpsi\psi(4160)$; $\phi_i~(i=2,3)$ is the interference phase as expressed in Eq.~(\ref{flatte})\footnote{The remaining parameters which are not listed here are displayed in the Tab.~\ref{tab:swave:fit2}.}}
    \label{tab:swave:fit}
    \renewcommand\arraystretch{1.2} 
    \begin{tabular*}{\textwidth}{@{\extracolsep{\fill}}c|cccccc}
       \hline
       \hline
               &   \multicolumn{2}{c}{Case I}  &\multicolumn{2}{c}{Case  II} & \multicolumn{2}{c}{Case III}     \\
               &  Solution \Rone & Solution \Rtwo & Solution \Rone & Solution \Rtwo  & Solution \Rone & Solution \Rtwo     \\
       \hline
       $\chi^2/$d.o.f.   & $100.1/86$              & $100.6/86$           & $97.6/86$              & $96.4/86$             & $99.2/86$             & $99.6/86$            \\
       $M_2$~(MeV)     & $6883.3\pm100.3$  & $6881.9\pm203.2$  & $6921.5\pm147.6$  & $6850.0\pm136.7$ & $6829.8\pm113.6$  & $6850.0\pm107.8$ \\
       $g_{21}$~(MeV)   & $338.8\pm25.8$     & $1029.1\pm91.1$     & $1000.7\pm19.3$    & $1006.6\pm56.7$   & $606.5\pm18.8$     & $1005.9\pm54.1$   \\
       $g_{22}$~(MeV)  & $110.9\pm123.0$     & $1644.1\pm244.4$   & $645.9\pm56.9$    & $1683.5\pm239.1$  & $259.6\pm71.6$     & $1661.9\pm245.5$ \\
       $\phi_2$~(rad)    & $0.7\pm1.6$           & $1.3\pm1.7$            & $3.9\pm0.9$         & $1.8\pm1.1$            & $2.7\pm1.6$          & $2.1\pm0.9$          \\
       $M_3$~(MeV)     & $7195.1\pm212.8$    & $7150.0\pm747.5$  & $7221.1\pm172.4$  & $7165.6\pm656.0$  & $7222.2\pm182.9$ & $7169.9\pm583.2$ \\
       $g_{31}$~(MeV)   & $68.5\pm12.4$        & $151.5\pm48.6$     & $120.0\pm31.0$      & $130.4\pm52.3$     & $110.0\pm32.2$     & $127.2\pm57.7$     \\
       $g_{32}$~(MeV)  & $94.1\pm92.9$        & $832.3\pm245.7$   & $0.0004\pm151.9$  & $774.5\pm262.8$   & $0.0003\pm155.4$ & $772.7\pm 251.7$  \\
       $\phi_3$~(rad)    & $5.5\pm0.9$           & $0.8\pm1.1$           & $4.4\pm1.9$          & $1.1\pm1.0$            & $5.2\pm1.4$          & $1.2\pm0.9$         \\
       \hline
       \hline
    \end{tabular*}
\end{table*}


One can use each set of the parameters  to determine whether the resonance structure studied in this paper is a confining state or a molecular state.  The definition of Riemann sheets for two channels is listed in Tab.~\ref{tab:sheet4}. 
The pole positions in  ${s}$  plane obtained by using parameters in Tab.~\ref{tab:swave:fit} for all cases are summarized in Tab.~\ref{tab:swave:pole}.
For Solution I, that pole positions of $X(7200)$  on the second and third sheets are equal in Case I and Case II indicates the $X(7200)$ state hardly couples to the $\jpsi \psi(4160)$ channel, whereas its pole positions in Case III manifests that it tends to be a confining state.
Furthermore, it is evident for this solution in each case that the co-existence of two poles near the $X(6900)$ threshold  indicates that it might be a confining state, for $S$-wave couplings.
For Solution II in each case, that only one pole is found on sheet II near the second threshold demonstrates that the two states tend to be molecular states.
Thus, in the case of assuming $X(6900)$ and $X(7200)$ being $J^{PC}=(0,2)^{++}$ and considering the  couple channels listed in Tab.~\ref{tab:Swave}, different conclusions with the goodness of the fits being almost equivalent are drawn. It is mainly caused by low statistics and unavailable information on other channels.
As a consequence,  it is impossible to distinguish whether the two states are confining states or molecular states under the current situation.
More experimental measurements in the couple channels, $X(6900)\to\jpsi \psi(3770)$, $\jpsi \psi_2(3823)$, $\jpsi \psi_3(3842)$, ~$\chi_{c0} \chi_{c1}$, ~and ~$X(7200)$ $\to$ $\jpsi\psi(4160)$, $\chi_{c0}$ $\chi_{c1}(3872)$, are therefore in urgent need to clarify their nature.

\begin{table}[htpb]
    \centering
    \caption{Definition of Riemann sheets ($i=2,3$).}
    \label{tab:sheet4}
    \begin{tabular*}{\columnwidth}{@{\extracolsep{\fill}}ccccc}
    \hline
    \hline\noalign{\smallskip}
         &\quad\quad I \quad \quad & \quad\quad II \quad  \quad  & \quad\quad III \quad \quad   & \quad\quad IV \quad \quad  \\
         \hline
       $\rho_{i1}$ & \quad + & \quad - & ~- & ~+\\
       $\rho_{i2}$ & \quad + & \quad  + & ~- & ~- \\
       \noalign{\smallskip}\hline
       \hline
    \end{tabular*}
\end{table}

\begin{table}[htpb]
    \centering
    \caption{Summary of pole positions which are obtained using central values of parameters for  $S-S$ couplings. Here, the symbol ``Sol." denotes ``Solution" throughout the analysis.}
    \label{tab:swave:pole}
    \renewcommand\arraystretch{1.2} 
    \begin{tabular*}{\columnwidth}{@{\extracolsep{\fill}}ccccc}
       \hline
       \hline\noalign{\smallskip}
                  & Case &  State    & Sheet II & Sheet III    \\
        \hline
        \multirow{6}{*}{ \threes Sol. I }       & \multirow{2}*{ I}       & $X(6900)$    & $6885.4-68.0 i$ & $6874.4-80.0 i$    \threes    \\
                                                                &                                      & $X(7200)$   & $7202.2-16.6 i$  & $7187.1-18.0 i$       \threes    \\
                                                                & \multirow{2}*{ II}       & $X(6900)$   & $6947.6-172.0 i$ & $6810.4-274.0 i$   \threes   \\
                                                                &                                      & $X(7200)$   & $7220.8-31.0 i$  & $7220.8-31.0 i$      \threes   \\
                                                                & \multirow{2}*{III}      & $X(6900)$   & $6845.2-117.0 i$ & $6789.2-138.0 i$    \threes   \\
                                                                &                                      & $X(7200)$   & $7221.9-28.0 i$  & $7221.9-28.0 i$      \threes  \\
        \hline
        \multirow{6}{*}{ \threes Sol. II }       & \multirow{2}*{ I}        & $X(6900)$   & $6937.9-97.0 i$  & $6527.3-323.0 i$   \threes    \\
                                                                &                                       & $X(7200)$   & $7210.7-27.5 i$   & $7037.3-47.5 i$     \threes   \\
                                                                & \multirow{2}*{II}        & $X(6900)$   & $6933.9-111.0 i$ & $6443.8-275. 0i$    \threes  \\
                                                                &                                       & $X(7200)$   & $7218.9-24.0 i$  & $7067.9-41.5 i$       \threes  \\
                                                                & \multirow{2}*{III}       & $X(6900)$   & $6933.3-113.0 i$ & $6452.3-275.0 i$    \threes  \\
                                                                 &                                      & $X(7200)$   & $7221.9-23.0 i$  & $7073.7-41.0 i$       \threes  \\
       \hline
       \hline
    \end{tabular*}
\end{table}

At last, we also test the situation that $X(6900)$ couples to $\jpsi\jpsi$ and $\jpsi\psi(2S)$, and a solution that favors $X(6900)$ as a confining state is found. 
Meanwhile, we can not find a good solution in favor of a molecular state interpretation of $X(6900)$.

\subsection{$P-P$ couplings}
\begin{figure}[htpb]
    \vskip 0.4cm
    \centering
    \subfigure[~Case I]{\includegraphics[scale=0.22]{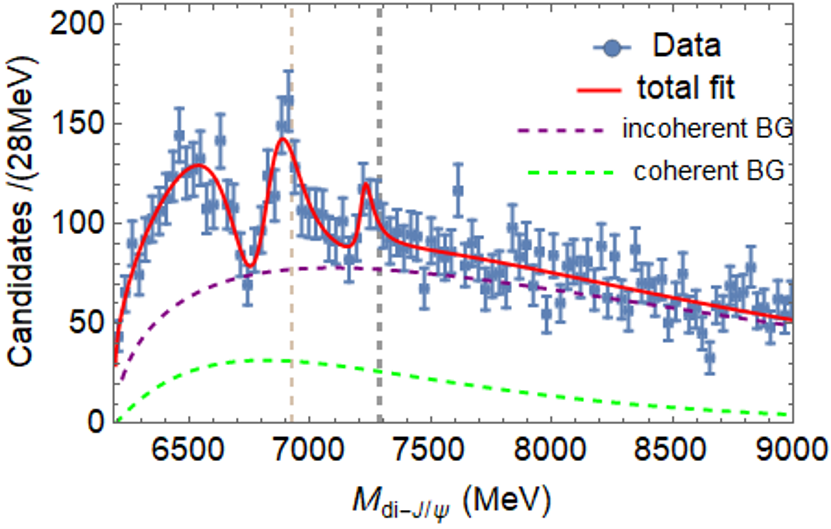}}
    \quad
    \subfigure[~Case II]{\includegraphics[scale=0.22]{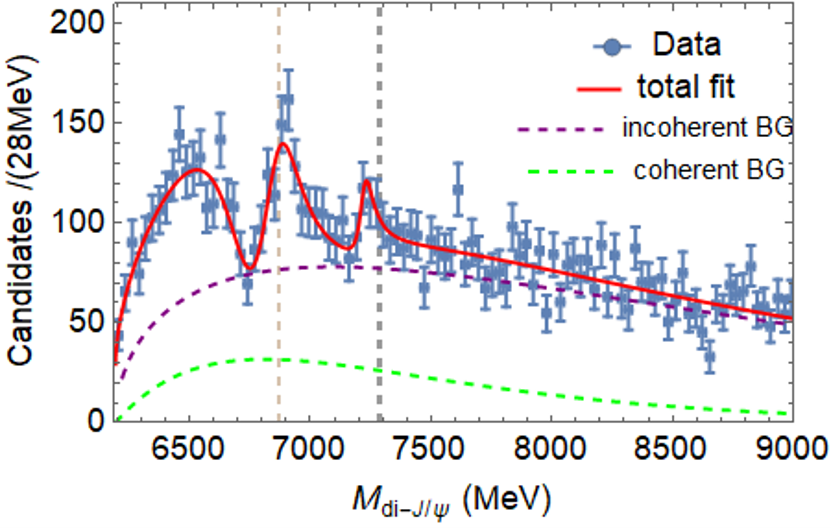}}
    \caption{(Color online) Fit projections for the $P-P$ couplings, where Fig.~(a)  shows $X(6900)$ decaying to $\jpsi \jpsi$ and $\chi_{c0} \chi_{c1}$, and $X(7200)$ to $\jpsi\jpsi$ and $\chi_{c0}\chi_{c1}(3872)$, and Fig.~(b)  illustrates $X(6900)\to\jpsi \jpsi$ and $\jpsi \psi(3770)$, $X(7200)\to\jpsi\jpsi$ and $\jpsi \psi(4160)$. Here, the descriptions of the components of the figures are similar to those of Fig.~\ref{fig:sscouple1}. }
    \label{fig:pp}
\end{figure}

With the quantum numbers of the $P$-wave  $\jpsi \jpsi$ pair being $(0,1,2)^{-+}$,
couple channel thresholds near to the two states are summarized in Tab.~\ref{tab:Pwave}.
From this table, the couplings can be divided into two cases:\\
\indent Case I: $X(6900)\to\jpsi \jpsi$, $\chi_{c0} \chi_{c1}$; $X(7200)\to$ $\jpsi\jpsi$, $\chi_{c0}\chi_{c1}(3872)$.\\
\indent Case II: ~$X(6900)\to\jpsi \jpsi$, $\jpsi \psi(3770)$; $X(7200)$ $\to\jpsi\jpsi$ and $\jpsi \psi(4160)$.  \\
By employing Eq.~(\ref{flatte}) with the threshold and barrier factors $n_{ij}(s)$ included, the fit projections are shown in Fig.~\ref{fig:pp}, and the corresponding numerical results are listed in Tab.~\ref{tab:pwave:fit}.
It can be seen that the parameterization with the $P$-wave coupling assumption can also meet the experimental data well with almost equivalent  the goodness of the fit in the $S$-wave couplings.
The pole positions in the complex $s$ plane are listed in Tab.~\ref{tab:pwave:pole}.
It should be stressed that the method adopted in this paper can not distinguish a $P$-wave confining state from a $P$-wave molecule, since they both contribute two pair of poles near the threshold.

\begin{table}[htpb]
    \centering
    \caption{Summary of numerical results for $P-P$ couplings, where $M_2$ and $M_3$ are mass of $X(6900)$ and $X(7200)$, respectively; $g_{21}$ and $g_{31}$ are the coupling constants of $X(6900)$ and $X(7200)$ decaying to $\jpsi \jpsi$, respectively;  $g_{22}$ is the coupling constant of $X(6900)\to\chi_{c0}\chi_{c1}$ and $\jpsi \psi(3770)$ in turn in the two cases, $g_{32}$ is the coupling constant of $X(7200)\to \jpsi\psi(4160)$; $\phi_i~(i=2,3)$ is the interference phase as expressed in Eq.~(\ref{flatte})\footnote{The remaining parameters which are not listed here are displayed in the Tab.~\ref{tab:pwave:fit2}.}}
    \label{tab:pwave:fit}
    \renewcommand\arraystretch{1.2} 
    \begin{tabular*}{\columnwidth}{@{\extracolsep{\fill}}ccc}
       \hline
       \hline
               &   Case  I  &  Case  II     \\
       \hline
       \threes$\chi^2/$d.o.f.   & $95.7/86$              & $95.6/86$               \\
       \threes$M_2$~(MeV)    & $6866.9\pm51.4$     & $6869.6\pm60.2$     \\
       \threes$g_{21}$~(MeV)  & $4679.6\pm128.6$   & $4679.2\pm136.3$        \\
       \threes$g_{22}$~(MeV) & $1010.1\pm1520.1$    & $2585.2\pm1828.7$      \\
       \threes$\phi_2$~(rad)   & $1.7\pm1.4$            & $1.7\pm1.6$               \\
       \threes$M_3$~(MeV)    & $7228.2\pm50.1$    & $7229.7\pm50.7$     \\
       \threes$g_{31}$~(MeV)  & $568.8\pm60.6$     & $569.3\pm59.8$         \\
       \threes$g_{32}$~(MeV) & $7105.6\pm5964.6$ & $8158.6\pm5610.5$      \\
       \threes$\phi_3$~(rad)   & $5.7\pm0.6$           & $ 5.7\pm0.6$              \\
       \hline
       \hline
    \end{tabular*}
\end{table}

\begin{table}[htpb]
    \centering
    \caption{Summary of pole positions which are obtained using central values of parameters for the $P-P$ couplings.}
    \label{tab:pwave:pole}
    \renewcommand\arraystretch{1.2} 
    \begin{tabular*}{\columnwidth}{@{\extracolsep{\fill}}cccc}
       \hline
       \hline\noalign{\smallskip}
             &   State     & Sheet II & Sheet III   \\
        \hline
        \multirow{2}{*}{Case I}   & $X(6900)$  & $6838.7-119.0 i$ & $6840.9-113.0 i$      \\
                                      & $X(7200)$ & $7220.8-31.0 i$  & $7232.5-23.0 i$  \\
        \hline
        \multirow{2}{*}{Caes II}  & $X(6900)$ & $6844.1-122.0 i$ & $6841.2-110.5 i$      \\
                                      & $X(7200)$  & $7221.4-32.0 i$  & $7234.4-22.0 i$      \\

       \hline
       \hline
    \end{tabular*}
\end{table}

\subsection{$S-P$ and $P-S$ couplings}
Owing to the limited statistics but multiple states, only two cases are considered in the analysis for the $S-P$ and $P-S$ couplings:\\
\indent Case I  ($S-P$): $S$-wave for $X(6900)\to\jpsi \jpsi$, $\jpsi \psi(3770)$; $P$-wave for $X(7200)$ $\to\jpsi\jpsi$ and $\jpsi \psi(4160)$.  \\
\indent Case II ($P-S$):  $P$-wave for $X(6900)\to\jpsi \jpsi$, $\jpsi \psi(3770)$; $S$-wave for $X(7200)$ $\to\jpsi\jpsi$ and $\jpsi \psi(4160)$.  \\
The following total amplitude is applicable for the two cases,
\begin{equation}\label{S-P}
    \begin{aligned}
      |\mathcal{M}|^2= &\Bigg| \sum_{i=1}^{2} \mathcal{M}_i + \mathcal{M}_{\rm NoR} \Bigg|^{2} + |\mathcal{M}_3|^{2} + \mathcal{B.G.}
    \end{aligned}
\end{equation}
By employing Eq.~(\ref{S-P}) with the respective threshold and barrier factors $n_{ij}(s)$ included, the fit projections are similar to Fig.~\ref{fig:sscouple1}.
Two sets of solutions with almost equivalent goodness of the fit are found in Case I.
Only one solution in favor of a molecular interpretation of $X(7200)$ is found in Case II.
The pole positions in the complex $s$ plane are summarized in Tab.~\ref{tab:s-p:pole}.
Using PCR, it may be concluded that Solution I favors that $X(6900)$ is a confining state and Solution II supports that it is a molecular bound state.

\begin{table}[htpb]
    \centering
    \caption{Summary of pole positions which are obtained using central values of parameters for  $S-P$ and $P-S$ couplings.}    
    \label{tab:s-p:pole}
    \renewcommand\arraystretch{1.2} 
    \begin{tabular*}{\columnwidth}{@{\extracolsep{\fill}}ccccc}
       \hline
       \hline\noalign{\smallskip}
                                         &      Sol.    &  State      & Sheet II & Sheet III    \\
        \hline
        \multirow{4}{*}{ Case I  }            & \multirow{2}*{ I}       & $X(6900)$    & $6901.0-32.6 i$ & $6884.4-61.7 i$        \\
                                                     &                                      & $X(7200)$   & $7196.2 -19.5 i$  & $7200.8-17.4 i$        \\
                                                      & \multirow{2}*{II}        & $X(6900)$   & $6894.8-65.3 i$     & $-$       \\
                                                      &                                       & $X(7200)$   & $7097.8-17.6 i$   & $7128.1-14.0 i$       \\
        \hline
        \multirow{2}{*}{  Case II }          &                                    & $X(6900)$   & $6900.5-14.5 i$ & $6900.3-15.2 i$     \\
                                                     &                                      & $X(7200)$   & $7362.2-67.9 i$  & $-$         \\
       \hline
       \hline
    \end{tabular*}
\end{table}

~\\

\section{Spectral density function sum rule}
In the case of $S$-waves, SDFSR can be utilized to provide insights into the nature of the $X(6900)$ state.
Ref.~\cite{Baru:2003qq} pointed out that the spectrum density function $\omega(E)$ near  threshold
can be calculated by using the non-relativistic $S$-wave Flatt\'{e} parameterization,
and the renormalization constant $\mathcal{Z}$ can be obtained,
which represents the probability of finding the confining particle in the continuous spectrum: the more the value of
$\mathcal{Z}$ tends to $1$, the more confining the state is. On the other hand, if $\mathcal{Z}$ tends to 0, the state tends to be molecular.

Using the similar form in Refs.~\cite{X3900,Baru:2003qq,Weinberg,Weinberg:1965zz,Kalashnikova:2009gt}, the spectrum density function of a near-threshold channel can be expressed as Eq.~(\ref{eq:spec}),
\begin{equation}
\label{eq:spec}
    \omega(E)=\frac{1}{2\pi}\frac{\tilde{g}\sqrt{2\mu E}\theta(E)+\tilde{\Gamma}_{0}}{\left|E-E_{f}+\frac{i}{2}\tilde{g}\sqrt{2\mu E}\theta(E)+\frac{i}{2}\tilde{\Gamma}_0\right|^2},
\end{equation}
where,  $E~(E_f)=\sqrt{s}~(M)-m_{th}$ is the energy difference between the center-of-mass energy (resonant state) and the open-channel threshold, $\mu$ the reduced mass of the two-body final states of the channel, $\theta$  the step function,
$\tilde{g}=2g/m_{th}$, the dimensionless coupling constant of the concerned coupling mode, and $\tilde{\Gamma_0}$ the constant partial width for the remaining couplings, which mainly contains the distant channels (the $\jpsi\jpsi$ process in this analysis).

\begin{table}[htbp]
    \centering
    \caption{Summary of $\mathcal{Z}$ values for the $S-S$ couplings of $X(6900)$. }
    \label{tab:zvalue}
    \renewcommand\arraystretch{1.2} 
    \begin{tabular*}{\columnwidth}{@{\extracolsep{\fill}}cccc}
       \hline
       \hline\noalign{\smallskip}
           &  Case   &  $[E_f-\Gamma,E_f+\Gamma]$ & $[E_f-2\Gamma,E_f+2\Gamma]$   \\
        \noalign{\smallskip}\hline\noalign{\smallskip}
        \multirow{3}{*}{ \threes Sol. I }   &  I  & 0.459   & 0.671       \\
                                                      & II  & 0.379   & 0.592     \\
                                                      & III & 0.468   & 0.681      \\
        \hline\noalign{\smallskip}
        \multirow{3}{*}{ \threes Sol. II}   &  I  & 0.184   & 0.344      \\
                                                      & II  & 0.243  & 0.418       \\
                                                      & III & 0.259   & 0.438      \\
       \noalign{\smallskip}\hline
       \hline
    \end{tabular*}
\end{table}

By integrating Eq.~(\ref{eq:spec}), the probability of finding an ``elementary" particle in the continuous spectrum can be obtained,
\begin{equation}
    \mathcal{Z}=\int_{E_{\rm min}}^{E_{\rm max}}\omega(E)dE.
\end{equation}
The integral interval takes $E_f$ as the central  value. It is pointed out in Ref.~\cite{Kalashnikova:2009gt} that the integration interval needs to cover the threshold of the couple channel.
Since the $X(7200)$ state is not of significance ($\sim3~\sigma$ reported in the LHCb experiment~\cite{6900}), only the $\mathcal{Z}$ values of $X(6900)$, as listed in Tab.~\ref{tab:Swave},  are  calculated.
Expanding $\omega(E)$ near the threshold of each channel, and bringing  $X(6900)$'s mass $M^{pole}_2$ and width $\Gamma^{pole}_2$ extracted from the second Riemann sheet in Tab.~\ref{tab:swave:pole}, one can obtain the corresponding $\mathcal{Z}$ value, where $\tilde{\Gamma_0}=\Gamma_{21}(M^{\rm pole}_2)$ (see also Eq.~(\ref{eq:Gamma})).
The numerical $\mathcal{Z}$ values are summarized in Tab.~\ref{tab:zvalue}, where the interval $[E_f-\Gamma,E_f+\Gamma]$ covers all thresholds of the calculated channels.
The $\mathcal{Z}$ values in Solution I are all slightly less than 50\% in this interval, but rapidly exceeds 50\% in larger integral intervals.
Hence $X(6900)$ may be considered as a confining state in Solution I.
For Solution II, the $\mathcal{Z}$ values are much smaller than 50\% in the interval $[E_f-\Gamma,E_f+\Gamma]$ , and also less than 50\% in the interval $[E_f-2\Gamma,E_f+2\Gamma]$.
This suggests that the $X(6900)$ state is more likely a molecular state in Solution II.
As a conclusion, the nature of $X(6900)$ is consistently drawn from both PCR and SDFSR, based on the current limited data.
Hence we are not able to distinguish whether it is a confining state or a molecular bound state.

\section{Conclusion}
\label{sec:4}
In this analysis, the channels $\jpsi \psi(3770)$, $\jpsi \psi_2(3823)$, $\jpsi \psi_3(3842)$, and $\chi_{c0} \chi_{c1}$  [$\jpsi\psi(4160)$ and $\chi_{c0}\chi_{c1}(3872)$] close to the threshold of $X(6900)$  [$X(7200)$] are selected to study their couplings. 
Fitting to the recent LHCb data by the Flatt\'{e}-like parameterization with the momentum-dependent partial widths,  
we found that the lowest state in the di-$\psi$ mass spectrum with the same quantum numbers as $X(6900)$ is essential to describe the extremely deep dip below 6800 MeV by destructive interference.
The amplitude poles in the complex $s$ plane are gained.
For the $S$-wave $\jpsi \jpsi$ couplings, PCR and SDFSR are imposed to determine whether the structures are confining states (bound by color force) or molecular states. The two approaches give consistent conclusions that both confining states and molecular states are possible, or it is unable to distinguish the nature of the two states, if only the di-$\jpsi$ experimental data with current statistics are available.
It is also argured in Ref.~\cite{YQMa} that the current experimental data are not enough to give a definitive conclusion on the nature of $X(6900)$.
In addition, the $X(6900)\to J/\psi \psi(2S)$ coupling with the threshold far away from  $X(6900)$'s mass is taken into account, and our result disfavors the $X(6900)$ structure as a $J/\psi \psi(2S)$ molecular state.
In the end, we are looking forward to more experimental data and more decay channels to clarify the nature of $X(6900)$  and  $X(7200)$, as well as determining their $J^{PC}$ quantum numbers.
Reasonably, we suggest that experiments measure  $X(6900)\to$ $\jpsi \psi(3770)$, $\jpsi \psi_2(3823)$, $\jpsi \psi_3(3842)$, and $\chi_{c0} \chi_{c1}$;  and $X(7200)\to$ $\jpsi$$\psi(4160)$, $\chi_{c0}$$\chi_{c1}(3872)$ decays, which are expected to be available in LHCb, Belle-II, CMS and other (future) experiments.

\section*{Acknowledgements}
This work is supported in part by National Nature Science Foundations
of China under Contract Number 11975028 and 10925522;
and China Postdoctoral Science Foundation under Contract Number 2020M680500.

\appendix
\section{Values of Parameters}
The remaining parameters which are not listed in Sec.~II are presented below. The parameter values for $S-S$ coupling and  $P-P$ coupling are displayed in Tab.~\ref{tab:swave:fit2} and Tab.~\ref{tab:pwave:fit2}, respectively. 
\begin{table*}[htpb]
    \centering
    \begin{tabular*}{\textwidth}{@{\extracolsep{\fill}}c|cccccc}
       \hline
       \hline
               &   \multicolumn{2}{c}{Case I}  &\multicolumn{2}{c}{Case  II} & \multicolumn{2}{c}{Case III}     \\
               &  Solution \Rone & Solution \Rtwo & Solution \Rone & Solution \Rtwo  & Solution \Rone & Solution \Rtwo     \\
       \hline
       $g_1$~(MeV$^2$)   & $(34.5\pm 3.5)\times 10^6$ & $(56.2\pm 5.6)\times 10^6$           &$(35.0\pm 3.6)\times 10^6$ &$(37.2\pm 4.3)\times 10^6$ & $(57.8\pm 4.0)\times 10^6$ & $(28.2\pm 3.8)\times 10^6$\\
       $\phi_1$~(rad)  & $3.6\pm 0.1$  & $3.6\pm 0.1$  & $5.8\pm 0.1$  & $3.8\pm 0.1$ & $4.8\pm 0.1$  & $3.9\pm 0.1$ \\
        $M_1$~(MeV)     & $6621.6\pm 60.0$ & $6726.8\pm 164.0$ & $6741.0\pm 59.1$ & $6683.3\pm 155.9$   & $6737.0\pm 78.5$     & $6671.0\pm 101.6$   \\
       $g_{11}$~(MeV)  & $540.9\pm 55.6$ & $719.7\pm 66.0$   & $296.4\pm 22.2$    & $543.3\pm 53.8$  & $412.1\pm 33.4$     & $450.0\pm 51.2$ \\
       $g_{2}$~(MeV$^2$)   & $(11.9\pm 0.8)\times 10^6$& $(40.2\pm 5.6)\times 10^6$     & $(71.3\pm 4.7)\times 10^6$&$(32.4\pm 3.3)\times 10^6$&$(35.0\pm 5.6)\times 10^6$ & $(29.0\pm 2.9)\times 10^6$ \\
       $g_{3}$~(MeV$^2$)  & $(2.3\pm 0.3)\times 10^6$ &$(2.0\pm 0.6)\times 10^6$    & $(1.8\pm 0.4)\times 10^6$  & $(1.6\pm 0.5)\times 10^6$  & $(1.6\pm 0.4)\times 10^6$ &$(1.6\pm 0.5)\times 10^6$   \\
       \hline
       \hline
    \end{tabular*}
    \caption{Parameter values for $S-S$ coupling. Other parameter values are listed in Tab.~\ref{tab:swave:fit}.}
    \label{tab:swave:fit2}
\end{table*}

\begin{table}[htpb]
    \centering
    \begin{tabular}{c|cc}
    \hline
    \hline
                   & Case I                   & Case II\\
    \hline
    $g_1$~(MeV$^2$)&$(41.7\pm 3.1)\times 10^6$&$(41.7\pm 3.1)\times 10^6$  \\
    $\phi_1$~(rad) &$4.1\pm 0.1$              &$4.0\pm 0.1$ \\
    $M_1$~(MeV)    &$6753.8\pm 60.7$          &$6748.7\pm 85.5$ \\
    $g_{11}$~(MeV) &$402.8\pm 33.7$           &$414.6\pm 34.1$ \\
    $g_2$~(MeV$^2$)&$(74.3\pm 8.9)\times 10^6$&$(72.2\pm 8.4)\times 10^6$ \\
    $g_3$~(MeV$^2$)&$(8.0\pm 0.6)\times 10^6$ &$(7.9\pm 0.6)\times 10^6$ \\
    \hline
    \hline
    \end{tabular}
    \caption{Parameter values for $P-P$ coupling. Other parameter values are listed in Tab.~\ref{tab:pwave:fit}.}
    \label{tab:pwave:fit2}
\end{table}

For $S-P$ coupling, where $X(6900)$ couples to $S$-wave di-$J/\psi$ and $J/\psi\psi(3770)$, and $X(7200)$ couples to $P$-wave di-$J/\psi$ and $J/\psi\psi(4160)$, the parameter values are listed in Tab.~\ref{tab:fit:sp:para}.

\begin{table}[htpb]
    \centering
    \begin{tabular}{c|cc}
    \hline
    \hline
                  & Sol.~I & Sol.~II  \\
   \hline
   $\chi^2/d.o.f.$     & 98.2/87                     & 103.3/87\\
   $g_1$~(MeV$^2$)      &$(19.9\pm 4.2)\times 10^{6}$ & $(41.8\pm 5.4)\times 10^6$\\
   $\phi_1$~(rad)       &$3.4\pm 0.1$& $3.3\pm 0.1$ \\
   $M_1$~(MeV)          &$6599.5\pm 24.4$& $6670.7\pm 33.9$\\
   $g_{11}$~(MeV)       &$462.4\pm 91.2$& $752.9\pm 91.7$\\
   $g_2$~(MeV$^2$)      &$(50.0\pm 5.5)\times 10^6$&$(25.0\pm 1.3)\times 10^6$\\
   $\phi_2$~(rad)       &$2.0\pm 0.5$ &$1.1\pm0.2$ \\
   $M_2$~(MeV)          &$6896.1\pm 27.4$&$6800.0\pm 28.7$ \\
   $g_{21}$~(MeV)       &$215.8\pm 48.2$&$1000.0\pm 41.0$ \\
   $g_{22}$~(MeV)       &$250.0\pm 117.9$& $2550.0\pm 587.5$\\
   $g_3$~(MeV$^2$)      &$(7.6\pm 1.6)\times 10^6$&$(7.4\pm 2.9)\times 10^6$ \\
   $M_3$~(MeV)          &$7199.0 \pm 102.9$& $7115.3\pm 263.8$\\
   $g_{31}$~(MeV)       &$400.0\pm 106.3$ & $400.0\pm 127.5$\\
   $g_{32}$~(MeV)       &$1463.7\pm 257.6$ & $3424.8\pm 351.1$\\
   \hline
   \hline
    \end{tabular}
    \caption{Parameter values for $S-P$ coupling.}
    \label{tab:fit:sp:para}
\end{table}

For $P-S$ coupling, where $X(6900)$ couples to $P$-wave di-$J/\psi$ and $J/\psi\psi(3770)$, and $X(7200)$ couples to $S$-wave di-$J/\psi$ and $J/\psi\psi(4160)$, the parameter values are shown in Tab.~\ref{tab:fit:ps:para}. 
\begin{table}[htpb]
    \centering
    \begin{tabular}{c|c}
    \hline
   \hline
   $\chi^2/d.o.f.$     & 97.23/87 \\
   $g_1$~(MeV$^2$)      &$(29.5\pm 2.8)\times 10^{6}$\\
   $\phi_1$~(rad)       &$4.4\pm 0.2$\\
   $M_1$~(MeV)          &$6742.8\pm 17.9$\\
   $g_{11}$~(MeV)       &$1000.0\pm 80.8$\\
   $g_2$~(MeV$^2$)      &$(20.0\pm 2.5)\times 10^6$\\
   $\phi_2$~(rad)       &$2.3\pm 0.5$\\
   $M_2$~(MeV)          &$6900.8\pm 25.9$\\
   $g_{21}$~(MeV)       &$550.0\pm 31.6$ \\
   $g_{22}$~(MeV)       &$250.0\pm 40.6$\\
   $g_3$~(MeV$^2$)      &$(31.2\pm 10.2)\times 10^6$\\
   $M_3$~(MeV)          &$7255.8 \pm 115.4$\\
   $g_{31}$~(MeV)       &$1377.9\pm 505.0$\\
   $g_{32}$~(MeV)       &$3999.7\pm 2352.5$\\
   \hline
   \hline
    \end{tabular}
    \caption{Parameter values for $P-S$ coupling.}
    \label{tab:fit:ps:para}
\end{table}

In addtion, the parameters for both coherent and incoherent background terms, which are gained from fitting to the mass spectrum without the signal amplitude, are fixed throughout the default fits, in order to reduce the uncertainty of the multiple interference. For the coherent background, $c_0= 20.9$ and $c_1=-6.9\times 10^{-4}$~(MeV$^{-1}$). For the incoherent background which takes the similar form as the coherent background, it has two parameters: $a_0=240.7$, $a_1=-4.5\times 10^{-4}$~(MeV$^{-1}$).  

%
%

\end{document}